\documentclass[10pt, aps,prd,twocolumn,showpacs,superscriptaddress,nofootinbib,preprintnumbers]{revtex4-2}

\usepackage{amsmath, graphicx}
\usepackage{xcolor}
\usepackage[colorlinks,linkcolor=blue,anchorcolor=blue,citecolor=blue,urlcolor=blue,]{hyperref}
\usepackage{orcidlink}
\usepackage{booktabs}

\begin{document}

\title{How large are curvature perturbations from slow first-order phase transitions? \\
A gauge-invariant analysis}


\author{Xiao Wang\,\orcidlink{0000-0003-2271-1340}}
\affiliation{School of Physics and Astronomy, Monash University, Melbourne 3800 Victoria, Australia}

\author{Csaba Bal\'azs\,\orcidlink{0000-0001-7154-1726}}
\affiliation{School of Physics and Astronomy, Monash University, Melbourne 3800 Victoria, Australia}

\author{Ran Ding\,\orcidlink{0000-0002-2959-3140}}
\email[Contact author:~]{dingran@mail.nankai.edu.cn}
\noaffiliation

\author{Chi Tian\,\orcidlink{0000-0002-5891-8573}}
\email[Contact author:~]{ctian@ahu.edu.cn}
\affiliation{School of Physics, Anhui University, 111 Jiulong Road, Hefei, Anhui, China 230601}


\date{\today}

\begin{abstract}
When strongly supercooled   cosmological first-order phase transitions (FOPTs) are sufficiently slow, super-horizon inhomogeneities can be generated. We compute these super-horizon curvature perturbations by employing a gauge-invariant, multi-fluid formalism. By resolving the gauge ambiguities inherent in conventional separate-universe simulations, we demonstrate that Primordial Black Holes are unlikely to be produced by these super-horizon inhomogeneities. We also derive a fitting formula for the resulting curvature perturbations and discuss potential observational constraints on FOPTs imposed by limits on primordial curvature perturbations and associated scalar-induced gravitational waves.
\end{abstract}


\maketitle

\section{Introduction}

As one of the most appealing new physical phenomena in the early universe, cosmological first-order phase transitions (FOPTs) have been extensively studied. Such intriguing processes may occur when some fundamental symmetries are broken as the Universe cools down after inflation. 
These phase transitions may not only explain the observed cosmological baryon asymmetry~\cite{Kuzmin:1985mm,Cohen:1993nk,Rubakov:1996vz,Riotto:1999yt,Morrissey:2012db}, but may also serve as a predominant source of the stochastic gravitational-wave background (SGWB)~\cite{Hindmarsh:2013xza,Caprini:2019egz,Athron:2023xlk,Tian:2024ysd,Wang:2024slx,Tian:2025zlo,Balazs:2025jwq}, for which preliminary evidence in the nanohertz band has been reported by recent pulsar timing array (PTA) observations~\cite{NANOGrav:2023gor, Xu:2023wog, EPTA:2023fyk, Reardon:2023gzh, Miles:2024seg}.

If the SGWB signal in recent PTA datasets was attributed to FOPTs in the early universe,
strongly supercooled FOPTs are favored~\cite{NANOGrav:2023hvm}, suggesting a possible domination of vacuum energy prior to these FOPTs. As a consequence, a short period of thermal inflation may occur. In addition to favoring strong FOPTs, the PTA data also indicates that these transitions should be slow, characterized by a relatively small inverse duration parameter $\beta/H_n \sim 10$, where $H_n$ represents the Hubble parameter during nucleation. 
As a result, the bubble nucleation timescale can be comparable to the Hubble time. Given the stochastic nature of the Poissonian nucleation process, this leads to distinct nucleation histories across causally disconnected regions. If vacuum energy is also dominant, these variations in nucleation history can introduce large fluctuations in the total energy density of these regions, ultimately generating super-horizon inhomogeneities~\cite{Liu:2022lvz,Franciolini:2025ztf}.

These large scale inhomogeneities not only induce curvature perturbations~\cite{Liu:2022lvz,Cai:2024nln} and are therefore subject to current or projected constraints~\cite{Liu:2022lvz, Fu:2025tml}, but, as suggested in Ref.~\cite{Lewicki:2024ghw,Liu:2021svg,Kawana:2022olo,Gouttenoire:2023naa,Cai:2024nln, Kanemura:2024pae}, can also exceed the primordial black hole (PBH) formation threshold and lead to the production of PBHs. Furthermore, according to Ref.~\cite{Lewicki:2024ghw}, when these inhomogeneous modes re-enter the horizon following thermal inflation, scalar-induced gravitational waves (SIGWs) can be generated. The amplitudes of these SIGWs are comparable to those of the conventional SGWB arising from bubble collisions or the radiation fluid, making them detectable by PTAs or other next-generation gravitational wave observatories.

These rich observational consequences are direct results of the super-horizon density contrasts induced by slow and strong FOPTs. However, as first pointed out by Ref.~\cite{Franciolini:2025ztf}, caution must be taken when evaluating the gravitational effects arising from these density contrasts due to potential gauge ambiguities. Such density contrasts typically originate from statistical fluctuations in the bubble nucleation history and are computed by separate universe simulations without accounting for gravitational effects. When determining the gravitational perturbations induced by these inhomogeneities, gauge ambiguities arise because both the density contrast and the metric perturbations are gauge dependent. 

In this study, we present a gauge-invariant framework to evaluate the gravitational consequences of large-scale inhomogeneities generated by a FOPT.
Our methodology is based on the conventional formalism for multi-fluid systems, which enables the computation of gauge-invariant quantities, such as comoving curvature perturbations, without relying on a gauge-specific density contrast. Similarly as in~\cite{Franciolini:2025ztf}, we find that the production of both PBHs and SIGWs is significantly suppressed compared to previous studies. Utilizing the derived fitting formula relating gauge-invariant comoving curvature perturbations to phase transition parameters, we analyze the observational implications arising from current and projected constraints on primordial curvature perturbations, as well as from potential SIGW signals in the PTA data.

\section{Theory of Cosmological First-order Phase transitions}

In the standard framework for FOPTs, the nucleation rate of true vacuum bubbles is given by
\begin{align}
    \Gamma = H_n^4 \exp{[ \beta t]},
\end{align}
where $H_n$ is the Hubble parameter at the nucleation time $t_n$ (which has been set to $t_n=0$) and $\beta$ describes the nucleation rate. 

Once true-vacuum bubbles are generated, during the expansions of the bubbles, vacuum energy is transferred into the energy of the radiation, whose background average follows the continuity equation
\begin{align}
\label{eq:cont}
    \dot{\rho}_r+ 4 H \rho_r = -\dot{\rho}_V,
\end{align}
where $\rho_r$ and $\rho_V$ are the background energy density of the radiation and vacuum, respectively. The vacuum energy density $\rho_V$ can be evaluated from the average false vacuum fraction $\overline{F}(t)$~\cite{Turner:1992tz,Wang:2020jrd} by $\rho_V = \overline{F}(t) \Delta V$, where $\overline{F}(t)$ starts from 1 and ends at 0 after the FOPT, and $\Delta V$ denotes the vacuum energy density difference between the true and false vacua. The quantity $\overline{F}(t)$ can be estimated by 
\begin{align}
\label{eq:Ft}
    \overline{F}(t) = \exp\left[ -\frac{4\pi}{3} \int_{-\infty}^{t} \mathrm{d}t' \, \Gamma(t') a(t')^3 R(t, t')^3 \right],
\end{align}
where $R(t',t)$ is the comoving radius of a bubble nucleated
at time $t'$. During the thermal inflation driven by supercooled FOPTs, at the percolation time, the comoving Hubble radius will reach a minimum value, corresponding to a largest comoving wave number $k_{\rm max}=a(T_{\mathrm{reh}})H(T_{\mathrm{reh}})$, with $T_{\mathrm{reh}}$ being the reheating temperature. After percolation, modes with wave number $k<k_{\rm max}$ re-enter the horizon.

As suggested in Refs.~\cite{Liu:2022lvz, Elor:2023xbz,Lewicki:2024ghw}, if a strongly super-cooled FOPT occurred with a low nucleation rate ($\beta$ is small), the Poisson nature of the bubble nucleation processes will induce significant variations in the nucleation history at super-Hubble scales. Since the energy density of the radiation fluid decreases as $a(t)^4$ while the vacuum energy remains constant, a delayed nucleation within a given Hubble patch generally leads to a larger-than-average total energy density, described by a positive density contrast $\delta \equiv \delta \rho / \rho$. This phenomenon can be quantified by simulating nucleation histories and solving for the distinct time-evolutions of the density fields across a large ensemble of independent super-Hubble patches, an approach analogous to the conventional “separate universe” picture~\cite{Wands:2000dp}. Ref~\cite{Lewicki:2024ghw} provides a powerful toolkit \texttt{DeltaPT} to effectively facilitate such simulations.

However, these separate universe bubble nucleation simulations, such as in $\texttt{DeltaPT}$, are usually performed without specifying a gauge. We therefore denote the density contrast derived from such simulations as $\delta_{\rm NG}$ (no gauge). As first highlighted in~\cite{Franciolini:2025ztf}, ambiguities arise when identifying this density contrast with its counterpart in gravitational perturbation theory, which in turn affects the subsequent estimates of the PBH abundance or the magnitude of curvature perturbations. The authors obtain their results by identifying $\delta_{\rm NG}$ to the spatial flat gauge and solving the linearized Einstein equations with neglecting gravitational backreaction. In this work, rather than associating the results of $\texttt{DeltaPT}$ or similar simulations with a particular gauge choice, we assume that the density contrasts $\delta_{\rm NG}$ can be approximated, at linear order, by those computed on a specific but unknown time-slicing. This assumption enables the application of gauge-invariant formalisms. Utilizing the established formalism for multiple interacting fluids, detailed in the following section, we compute the curvature perturbations in a gauge-invariant manner.

\section{Large-scale curvature perturbation for multiple interacting fluids }

In this section, we outline the formalism for computing the curvature perturbation in the presence of multiple interacting fluids. For a comprehensive treatment, we refer the reader to Refs.~\cite{Malik:2002jb, Malik:2004tf}.

\subsection{Background equations}

Friedmann's equations and the corresponding time evolution of the Hubble parameter are
\begin{align}
H^2 &= \frac{8\pi G}{3}\rho, \\
\dot{H} &= -4\pi G(\rho + P) .
\end{align}
If the universe consists of multiple component fluids which satisfy
\begin{equation}
\sum_{\alpha} \rho_{\alpha} = \rho, \qquad \sum_{\alpha} P_{\alpha} = P,
\end{equation}
the continuity equation for each component is given by
\begin{equation}
\dot{\rho}_{\alpha} = -3H (\rho_{\alpha} + P_{\alpha}) + Q_{\alpha}, 
\end{equation}
where $Q_{\alpha}$ represents the energy transfer rate to the $\alpha$-fluid due to the interaction between different fluid components. Conservation of total energy implies that 
\begin{equation}
    \sum_{\alpha} Q_{\alpha} = 0.
\end{equation}
As an example, for a universe consisting of two primary components, radiation and vacuum, the continuity equation for the radiation component (Eq.~\ref{eq:cont}) is recovered by setting $\alpha=r$, $Q_r = -\dot{\overline{F}}(t)\Delta V$, and using the radiation equation of state, $P_r=\rho_r/3$.

\subsection{Perturbed equations}

In standard linear perturbation theory, the most general form of the line element for the spatially-flat FLRW background and scalar metric perturbations is described by~\cite{Bardeen:1980kt,Mukhanov:1990me}
\begin{widetext}
\begin{equation}
    ds^2 = -(1 + 2\phi)dt^2 + 2a \partial_i Bdtdx^i + a^2\big[(1 - 2\psi)\delta_{ij} + 2\partial_i \partial_j E \big]dx^idx^j,
\end{equation}
\end{widetext}
where $\psi$ and $\phi$ represent the gauge-dependent curvature perturbation and the lapse function, respectively.

With this perturbed metric, the gauge-invariant definition of the total curvature perturbations can be defined as 
\begin{align}
    \zeta \equiv  - \psi - H \frac{\delta \rho}{\dot{\rho}},
\end{align}
which coincides with the curvature perturbation $\psi$ for a uniform total-density slicing ($\delta \rho =0$). We can also define another gauge-invariant curvature perturbation $\mathcal{R}$ as 
\begin{align}
    \mathcal{R}\equiv  \psi - H a (v+B),
\end{align}
where $v$ is  the scalar velocity potential. This is the so-called comoving curvature perturbation, coinciding with $\psi$ on hypersurfaces orthogonal to worldlines comoving with the fluid.
These two gauge-invariant equations have the relation~\cite{Malik:2004tf}
\begin{align}
    \frac{k^2}{a^2}\Psi =  3 \dot{H} (\mathcal{R} + \zeta),
\end{align}
where another gauge-invariant curvature perturbation $\Psi \equiv \psi + H\sigma_s$ is introduced and $\sigma_s \equiv a^2 \dot{E} - aB$ is the scalar shear. Therefore, at super-Horizon scales ($k^2 / (aH)^2 \ll 1$), $R\approx -\zeta$.

From the perturbed continuity equations at super-horizon scales (see refs.~\cite{Wands:2000dp,Malik:2002jb,Malik:2004tf} for more details), the time evolution for the total density curvature perturbation is
\begin{align}
\label{eq:zeta_dot}
    \dot{\zeta} = - \frac{H}{\rho + P} \delta P_{\rm nad},
\end{align}
where the total non-adiabatic pressure perturbation $\delta P_{\rm nad}$ can be split into two parts,
\begin{align}
    \delta P_{\rm nad} \equiv \delta P_{\rm intr} + \delta P_{\rm rel}.
\end{align}
The first part represents the intrinsic entropy perturbation from each fluid
\begin{align}
    \delta P_{\rm intr} = \sum_{\alpha} \delta P_{\rm intr, \alpha},
\end{align}
where intrinsic non-adiabatic pressure perturbation of each fluid is given by 
\begin{align}
\label{eq:P_intr}
    \delta P_{\rm intr, \alpha} \equiv \delta P_{\alpha} - c_{\alpha}^2 \delta \rho_{\alpha}.
\end{align}
Here, the adiabatic sound speed of $\alpha$-fluid is defined as $c_{\alpha}^2\equiv \partial{P}_{\alpha}/ \partial{\rho}_{\alpha}$. By this definition, for any fluid with a definite equation of state of the form $P_{\alpha} = P_{\alpha}(\rho_{\alpha})$, the intrinsic non-adiabatic pressure perturbation, $\delta P_{\rm intr, \alpha}$, must vanish.

The second relative part of the non-adiabatic pressure perturbation can be computed by 
\begin{align}
\label{eq:P_rel}
    \delta P_{\text{rel}} = -\frac{1}{6H\dot{\rho}} \sum_{\alpha,\beta} \dot{\rho}_\alpha \dot{\rho}_\beta \left(c^2_\alpha - c^2_\beta\right) S_{\alpha\beta},
\end{align}
where the relative entropy perturbation $S_{\alpha \beta}$ is defined as 
\begin{align}
\label{eq:S_ab}
    S_{\alpha \beta}\equiv 3 (\zeta_{\alpha} - \zeta_{\beta}) = -3 H \left(\frac{\delta \rho_{\alpha}}{\dot{\rho}_{\alpha}} - \frac{\delta \rho_{\beta}}{\dot{\rho}_{\beta}}\right) .
\end{align}

\section{Large-scale curvature perturbations from FOPTs}

\subsection{Methodology}

At large scales, the dynamics of FOPTs can be effectively described by a system of multiple interacting fluids consisting of two primary components: radiation and vacuum.  We therefore adopt this gauge-invariant formalisms and plug in both components to compute the comoving curvature perturbations from FOPTs. Firstly, we use \texttt{DeltaPT} to simulate a large-amount of bubble nucleation histories at various super-Hubble scales and nucleation rates $\beta/H$. For each super-horizon wave number $k<k_{\rm max}$, we extract $50,000$ different realizations of the nucleation history in a box of size $4\pi k^{-3}/3$, represented by $50,000$ different functions of $F(t)$. Then, the same amount of different $\delta \rho_V(t) = (F(t) - \overline{F}(t))\Delta V$ are computed, where $F(t)$ is the false vacuum fraction computed for each causally disconnected patch. The corresponding $\rho_r(t)$ can then be obtained from the continuity equation Eq.~\eqref{eq:cont}.

Once the time evolution of $\delta \rho_V$ and $\delta \rho_r$ is determined, for each scale $k$ and nucleation rate $\beta/H$, we integrate Eq.~\eqref{eq:zeta_dot} up to $t_k$, the time at which the scale $k$ re-enters the horizon. In calculating $\delta P_{\rm nad}$, the intrinsic non-adiabatic perturbation $\delta P_{\rm intr}$ defined in Eq.~\eqref{eq:P_intr} vanishes, as the sound speeds of both fluids are assumed to be constant ($c_r^2 = 1/3$ and $c_V^2 = -1$), suggesting a definite equation of state. For the remaining relative component $\delta P_{\rm rel}$, we utilize Eq.~\eqref{eq:P_rel} and Eq.~\eqref{eq:S_ab} to perform the calculation.

It should be noted that $S_{V r}$ becomes divergent when $\dot{\rho}_r = 0$, as indicated by Eq.~\eqref{eq:S_ab}. This is because $\rho_r$ initially increases immediately after the onset of the FOPT and subsequently decreases around the percolation time due to cosmic expansion. However, when $S_{V r}$ is combined with the $\dot{\rho}_V\dot{\rho}_r$ term in Eq.~\eqref{eq:P_rel}, the divergence is canceled, resulting in a well-behaved integration. After obtaining $\zeta$ for different Hubble patches, the comoving curvature perturbations can be computed using the approximation at super-horizon scales $\mathcal{R}\approx -\zeta$. 

\subsection{Results}

\begin{figure}
    \centering
    \includegraphics[width=0.95\linewidth]{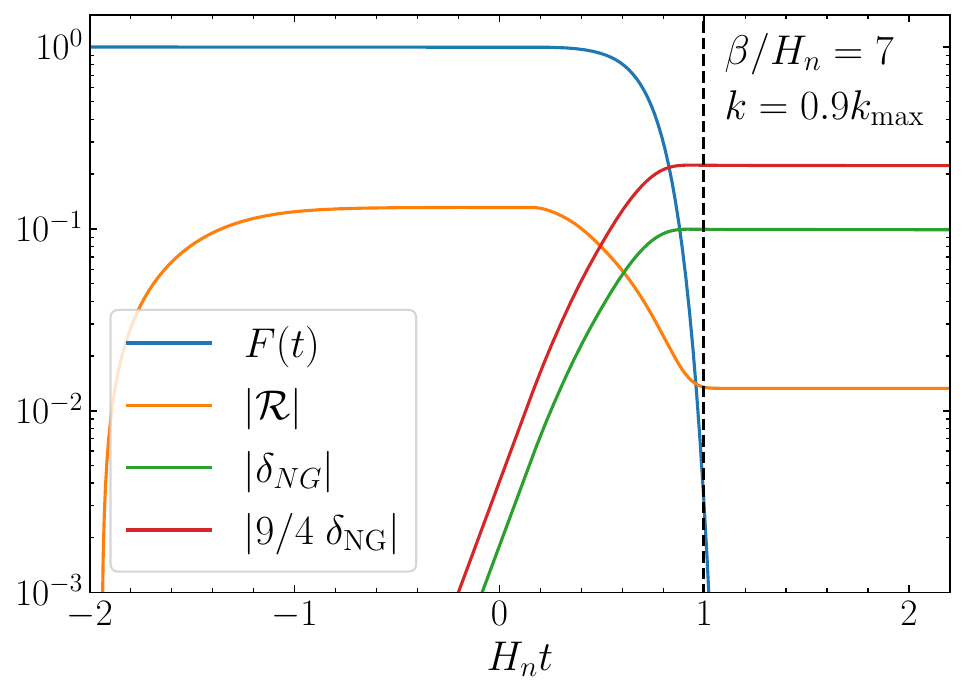}
    \caption{Time evolution of the absolute value of the false vacuum fraction $F(t)$, the comoving curvature perturbation $\mathcal{R}$, and the density contrast $\delta_{\rm NG}$ of an independently evolving volume $4\pi k^{-3}/3$, where $k=0.9 k_{\rm max}$ and $\beta / H_n =7$. The value of $ 9/4\; \delta_{\rm NG} $ is given for reference. The dashed line represents the time when the $k=0.9 k_{\rm max}$  mode re-enters the horizon.}
    \label{fig:0}
\end{figure}

We show the time evolution of the averaged comoving curvature perturbation $\mathcal{R}$ and density contrast $\delta$ in an independently evolving box of size $4\pi k^{-3}/3$ in Fig.~\ref{fig:0}, where $k=0.9 k_{\rm max}$. It is clear that there is no definite relation between $\delta_{\rm NG}$ and $\mathcal{R}$, and that the approximate relation for super-horizon perturbations of the radiation fluid $\mathcal{R}\approx -(9/4)\,\delta_C$~\cite{Green:2004wb} for comoving density contrast $\delta_C$ breaks down, suggesting that $\delta$ can not be approximately identified  as the value in the comoving gauge.

\begin{figure}
    \centering
    \includegraphics[width=0.98\linewidth]{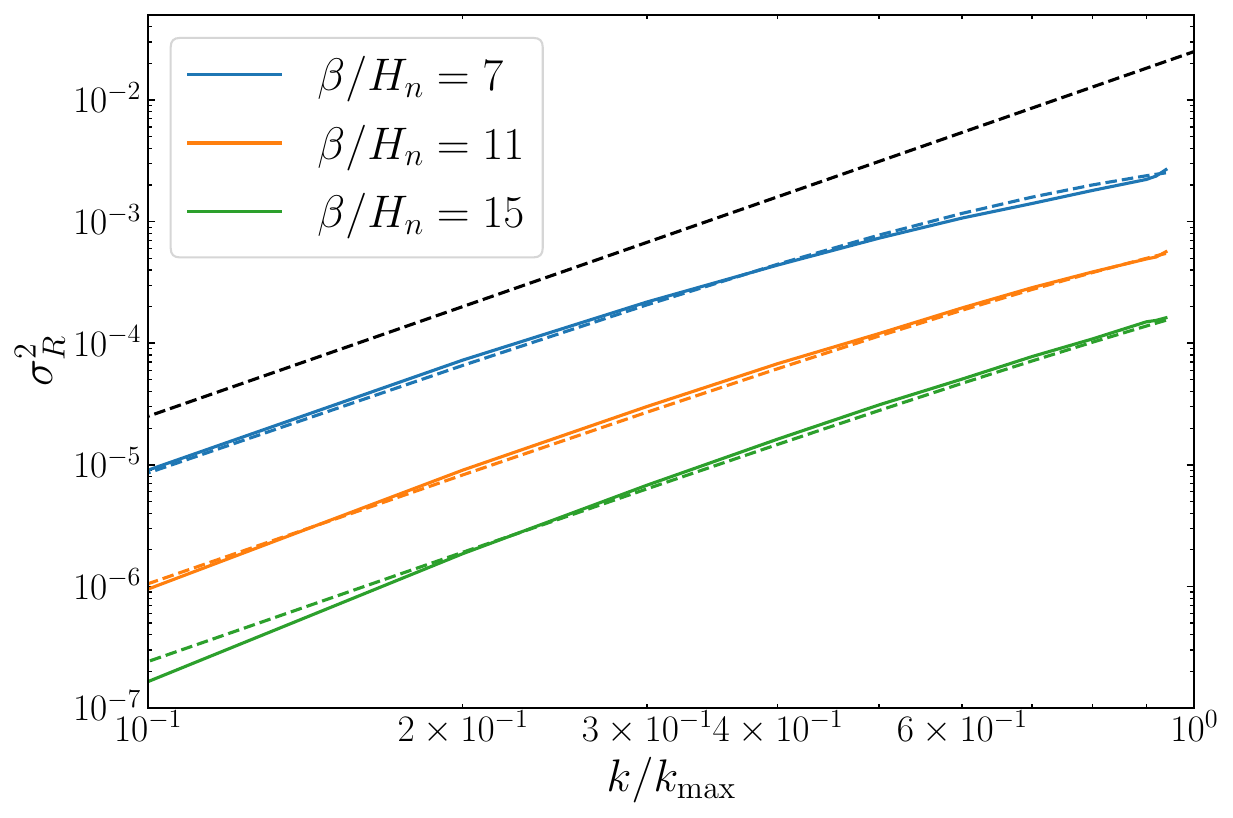}
    \includegraphics[width=0.98\linewidth]{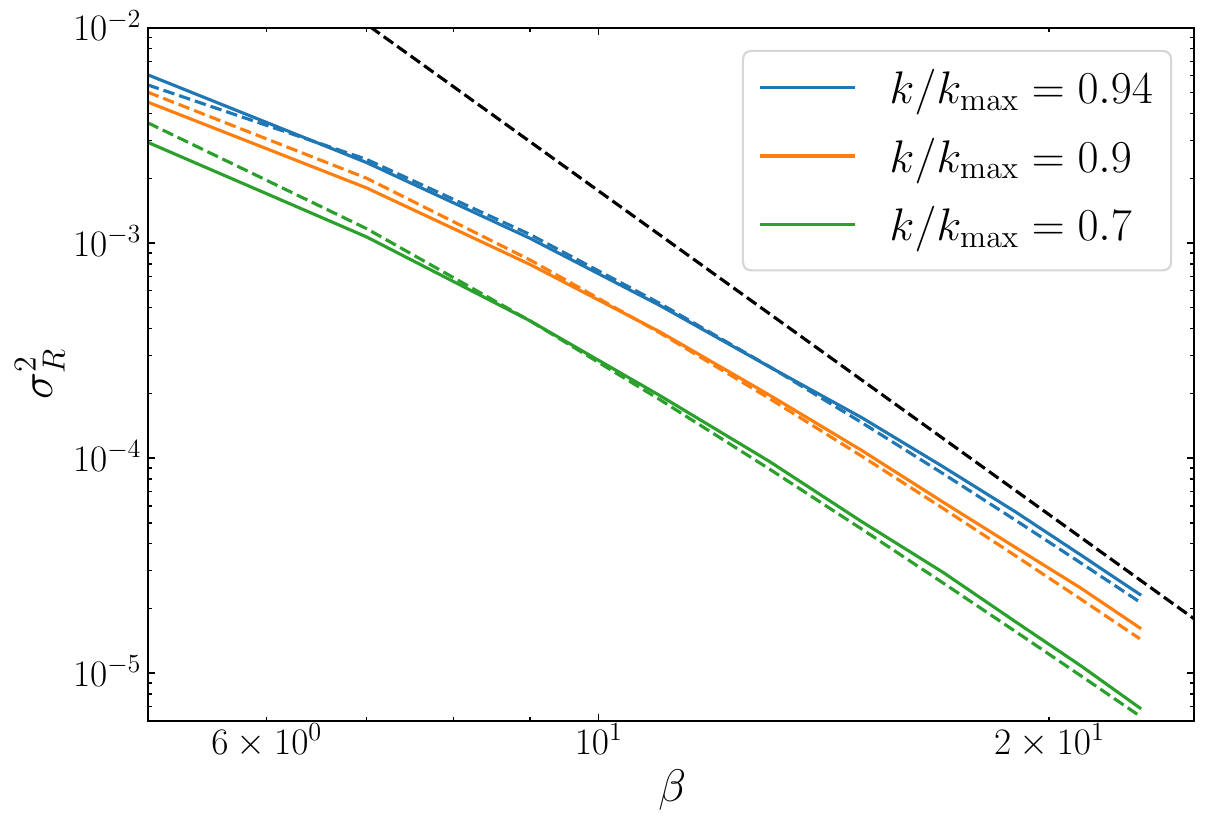}
    \caption{Variance of the comoving curvature perturbations $\sigma^2_{\mathcal{R}}$ as a function of $\beta$ and $k/k_{\rm max}$. The dashed curves represent the corresponding fits, while the dashed black lines indicate the $\beta^{-5}$ and $k^3$ reference power laws.}
    \label{fig:1}
\end{figure}

We also present in Fig.~\ref{fig:1} the variance of the resulting comoving curvature perturbations for various values of $k/k_{\rm max}$ and $\beta / H_n$. The magnitude of the variance is consistent with that reported in Ref.~\cite{Franciolini:2025ztf} and is significantly smaller than that found in~\cite{Lewicki:2024ghw}. The variance also exhibits the expected asymptotic behavior (see~\cite{Liu:2022lvz} for details), scaling as $k^3$ at small $k$ and as $\beta^{-5}$ at large $\beta$. However, for large $\beta$, small deviations from the $k^3$ scaling are observed, as shown in the upper panel of Fig.~\ref{fig:1}. These deviations are likely attributable to the limited number of bubbles employed in the simulations.

\begin{figure}
    \centering
    \includegraphics[width=0.92
    \linewidth]{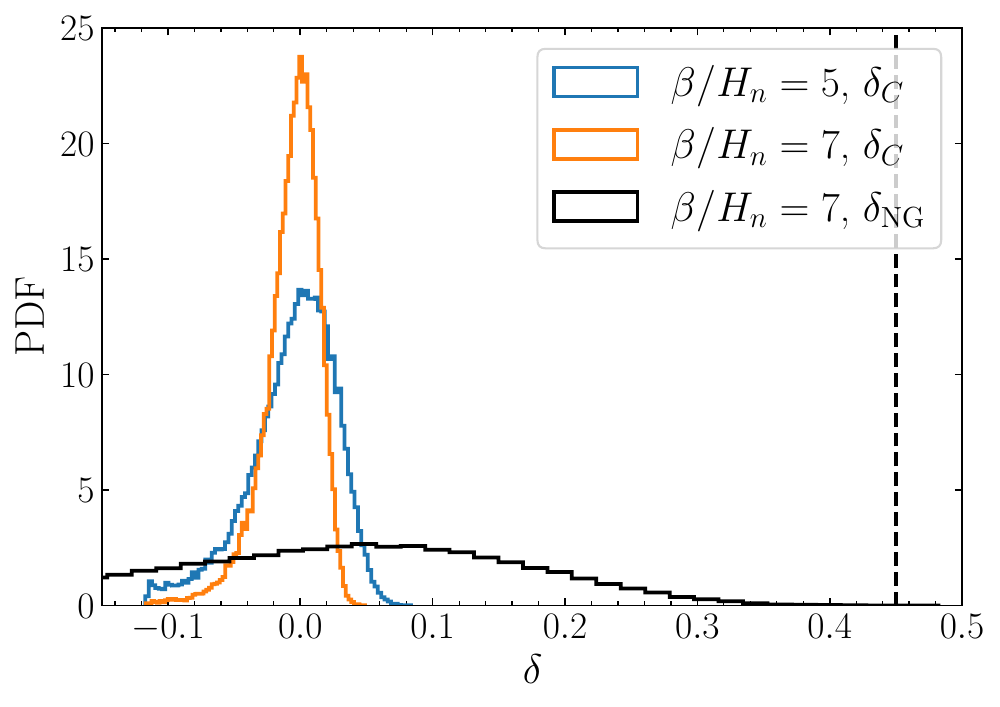}
    \caption{Distributions of density contrast of independently evolving volumes $4\pi k^{-3}/3$, where $k=0.9 k_{\rm max}$. The blue and orange envelopes represent $\delta_C$, the density contrast in the comoving-gauge. The black envelope represents the density contrast computed from separate universe simulations ($\delta_{\rm NG}$). The PBH formation threshold $\delta_{\rm crit} \approx 0.45$ is marked as a vertical dashed line.}
    \label{fig:2}
\end{figure}

Additionally, employing the approximated relation $\mathcal{R}\approx -(9/4) \delta_C$ on super-horizon scales, we estimate the probability distribution of the comoving density contrast, $\delta_C$, a key quantity for quantifying the PBH abundance. The results for $k=0.9k_{\rm kmax}$ are presented in Fig.~\ref{fig:2}. PBHs are believed to form from the positive tail of the $\delta_C$ distribution exceeding a critical threshold, $\delta_{\rm crit}$. Determining the precise value of $\delta_{\rm crit}$ remains an active area of research, due to its potential complex dependence on the shape of the perturbation~\cite{Escriva:2021aeh,Musco:2018rwt,Escriva:2019phb}, or requires careful evaluation of bubble wall structure and nonlinear collapse dynamics within the supercooled FOPT scenario~\cite{Ning:2026nfs}. As an illustration, we adopt the frequently quoted value $\delta_{\rm crit}\approx0.45$ in Fig.~\ref{fig:2}, which is derived from numerical simulations of PBH formation from curvature perturbations re-entering the horizon during radiation domination. A comparison of the $\delta_C$ distribution with this formation threshold indicates that PBHs are unlikely to be produced from the curvature perturbations generated by these FOPTs.

To accommodate the broader interests of phenomenological studies, we provide a fitting formula for $P_{\mathcal{R}}$ derived from a sampling of $\beta/H_n$ varying from 5 to 23 and $k/k_{\rm max}$ ranging from 0.1 to 0.94. We first fit $\sigma^2_{\mathcal{R}}$ (see the fitting results in Fig.~\ref{fig:1}), ensuring that the employed formula exhibits the correct asymptotic behavior. Using the approximate relation   
\begin{align}
\sigma^2_{\mathcal{R}} =\int \mathrm{d\,ln}k\,W^2(kR)P_{\mathcal{R}}(k) 
\simeq \frac{3\pi}{2}P_{\mathcal{R}}(k=R^{-1})\,,
\end{align}
which holds for the top-hat window function $W(x)=3(\sin{x}-x\cos{x})/x^3$ and assumes $P_{\mathcal{R}}\propto k^3$ on super-horizon scales, we derive the fitting formula for $P_{\mathcal{R}}$ as
\begin{align}
    P_{\mathcal{R}}(k) &= 0.0038\left(\frac{\alpha}{1+\alpha}\right)^2 \frac{(k/k_{\max})^3(\beta / H_n)^{-3}}{\left[0.043 + (k/k_{\max})^2(\beta / H_n)^{-2}\right]^3} \nonumber\\
&\quad \times
    \frac{1}{(1.77 + \beta / H_n)^2} \Theta(k_{\rm max} - k)\,.
    \label{eq:pr}
\end{align}
Where $\Theta$ is the heaviside step function, and $\alpha\approx \Delta V / \rho_r$ represents the phase transition strength. Given the limited accuracy of extrapolation, the fitting formula is recommended for use when $\beta / H_n \gtrsim 4$. The peak wave number $k_{\rm max}$ and the reheating temperature $T_{\rm reh}$ are related by
\begin{align}
k_{\max} &= \frac{\pi\, T_0}{3\sqrt{10}}\frac{T_{\rm{reh}}}{M_{\rm pl}}\left(\frac{g_s(T_{\rm{reh}})}{g_s(T_0)}\right)^{-1/3}g^{1/2}_\rho(T_{\rm{reh}}) \nonumber\\
&=1.61\times10^7\,{\rm{Mpc}}^{-1}\left(\frac{g_s(T_{\rm{reh}})}{68.74}\right)^{-1/3} 
\nonumber\\
&\quad \times \left(\frac{g_\rho(T_{\rm{reh}})}{69.76}\right)^{1/2}\left(\frac{T_{\rm{reh}}}{\rm{GeV}}\right) \,.
\end{align}
In above equation, $T_0=2.35\times10^{-13}$ GeV is the present-day CMB temperature~\cite{Fixsen:2009ug}, and $g_\rho$ and $g_s$ denote the effective degrees of freedom for the energy density and entropy density, where we adopt the values from Ref.~\cite{Saikawa:2020swg}.


\section{Observational consequences}

While inhomogeneities arising from FOPTs may be insufficient to generate PBHs, direct or indirect limits on primordial curvature perturbations from various astrophysical and cosmological probes can be employed to constrain FOPTs. Furthermore, when the corresponding modes re-enter the horizon following thermal inflation, SIGWs can arise as a direct consequence of curvature perturbations. In this section, we briefly discuss potential constraints on $P_{\mathcal{R}}(k)$ and the implications for FOPT parameters in the presence of SIGWs based on PTA data.

\subsection{Observational constraints on $P_{\mathcal{R}}$}

The primordial curvature power spectrum, $P_{\mathcal{R}}(k)$, produced by FOPTs is subject to various constraints. On large scales with modes $k \lesssim 3\ \mathrm{Mpc}^{-1}$, CMB anisotropy observations~\cite{Planck:2018vyg} and Lyman-$\alpha$ measurements~\cite{Bird:2010mp} restrict the spectrum to $P_{\mathcal{R}}\simeq \mathcal{O}(10^{-9})$. Enhancements of $P_{\mathcal{R}}$ on small scales can generate observable signals via CMB spectral distortions and SIGWs. The Far Infrared Absolute Spectrophotometer (FIRAS) has constrained the primordial power spectrum in the range of $10\ \mathrm{Mpc}^{-1} \lesssim k \lesssim 10^5\ \mathrm{Mpc}^{-1}$ through spectral observations of the CMB~\cite{Chluba:2012we}. Pulsar timing arrays and the upcoming Square Kilometer Array (SKA) telescope provide current limits and projected constraints for larger modes in the range $10^5\ \mathrm{Mpc}^{-1} \lesssim k \lesssim 10^{10}\ \mathrm{Mpc}^{-1}$~\cite{Byrnes:2018txb}. The next generation of gravitational wave observatories such as LISA will extend the sensitivity to even larger modes, covering the range $10^{10}\ \mathrm{Mpc}^{-1} \lesssim k \lesssim 10^{15}\ \mathrm{Mpc}^{-1}$~\cite{Inomata:2018epa}. The formation
of small dark matter structures, known as ultra-compact mini-halos (UCMHs), serves as an indirect probe of the primordial power spectrum. Assuming dark matter consists of weakly interacting massive particles that can pair-annihilate into standard model particles~\cite{Cirelli:2024ssz}, the presence of UCMHs can be effectively traced through $\gamma$-ray and neutrino observations~\cite{Scott:2009tu,Josan:2010vn,Bringmann:2011ut,Yang:2013dsa,Nakama:2017qac,Delos:2018ueo,Gouttenoire:2025wxc}, as well as imprints on CMB anisotropies resulting from extra energy injection~\cite{FrancoAbellan:2023sby}. 

\begin{figure*}[htbp]
    \centering
    \includegraphics[width=0.78\linewidth]{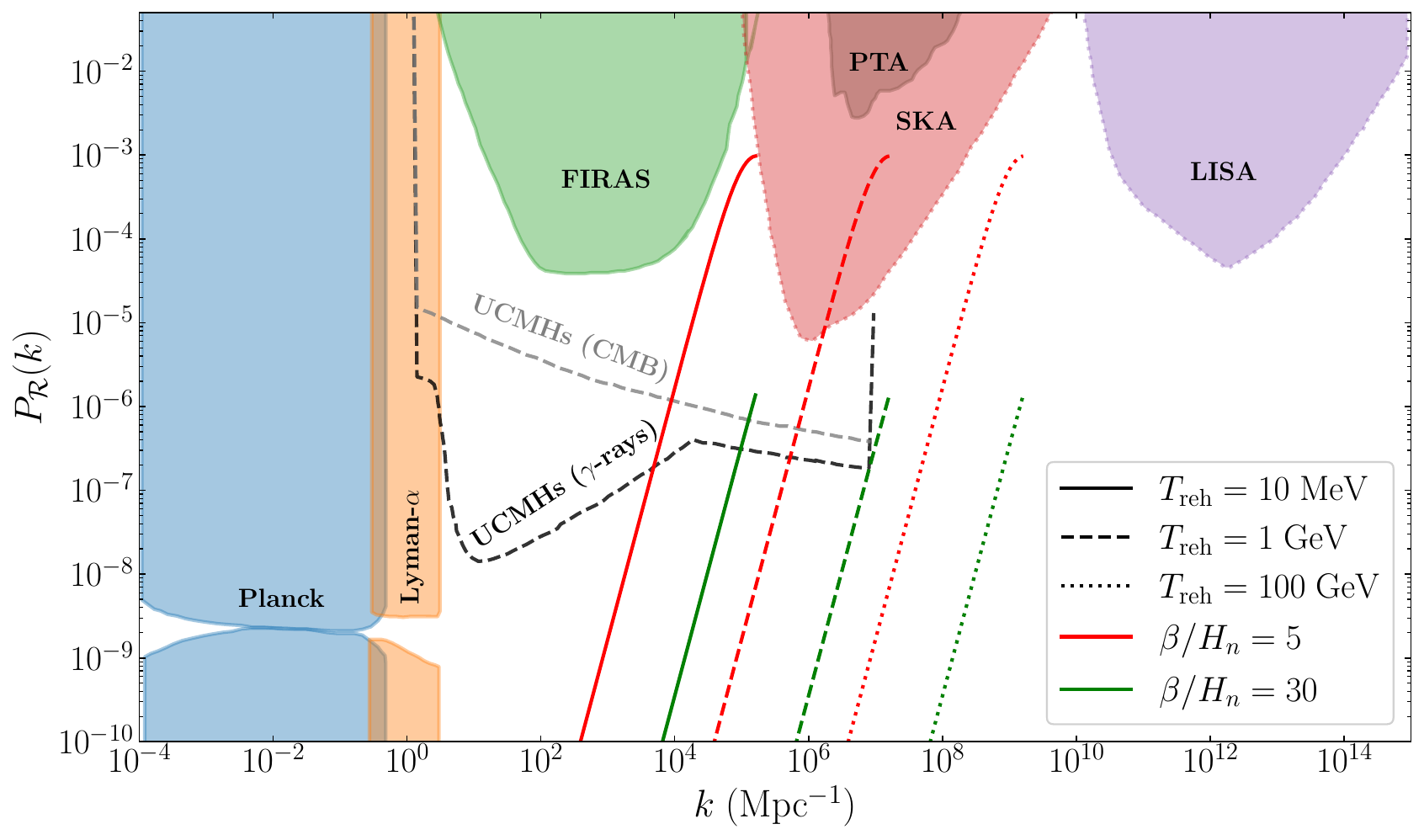}
    \caption{Various benchmark spectra  $P_{\mathcal{R}}(k)$ for $\alpha=10$ with varying $\beta/H_{n}$ and $T_{\rm reh}$. Also displayed is a compilation of existing constraints on the primordial power spectrum $P_{\mathcal{R}}(k)$ (solid-outlined filled regions), sensitivity forecasts for next-generation observatories (dotted-outlined filled regions), and UCMH-based constraints (dashed lines). The non-UCMH constraints are derived from CMB anisotropies~\cite{Planck:2018vyg} (light blue), Lyman-$\alpha$ observations~\cite{Bird:2010mp} (orange), CMB spectral distortions~\cite{Chluba:2012we} (green), Pulsar Timing Arrays~\cite{Byrnes:2018txb} (red), as well as SKA and LISA forecasts~\cite{Inomata:2018epa} (pink and purple). The UCMH-based constraints, which assume a monochromatic power-spectrum enhancement and that WIMP dark matter annihilates into into $b\bar{b}$ with a mass $m_{\rm DM}=1$ TeV and $s$-wave thermal relic cross section $\langle \sigma v \rangle_0=3 \times 10^{-26} \mathrm{cm}^{3}\mathrm{s}^{-1}$, are obtained from diffuse $\gamma$-ray searches~\cite{Delos:2018ueo} (black dashed line) and limits on extra energy injection into the CMB~\cite{FrancoAbellan:2023sby} (gray dashed line). }
    \label{fig:cons}
\end{figure*}

\begin{figure}[htbp]
    \centering
    \includegraphics[width=0.95\linewidth]{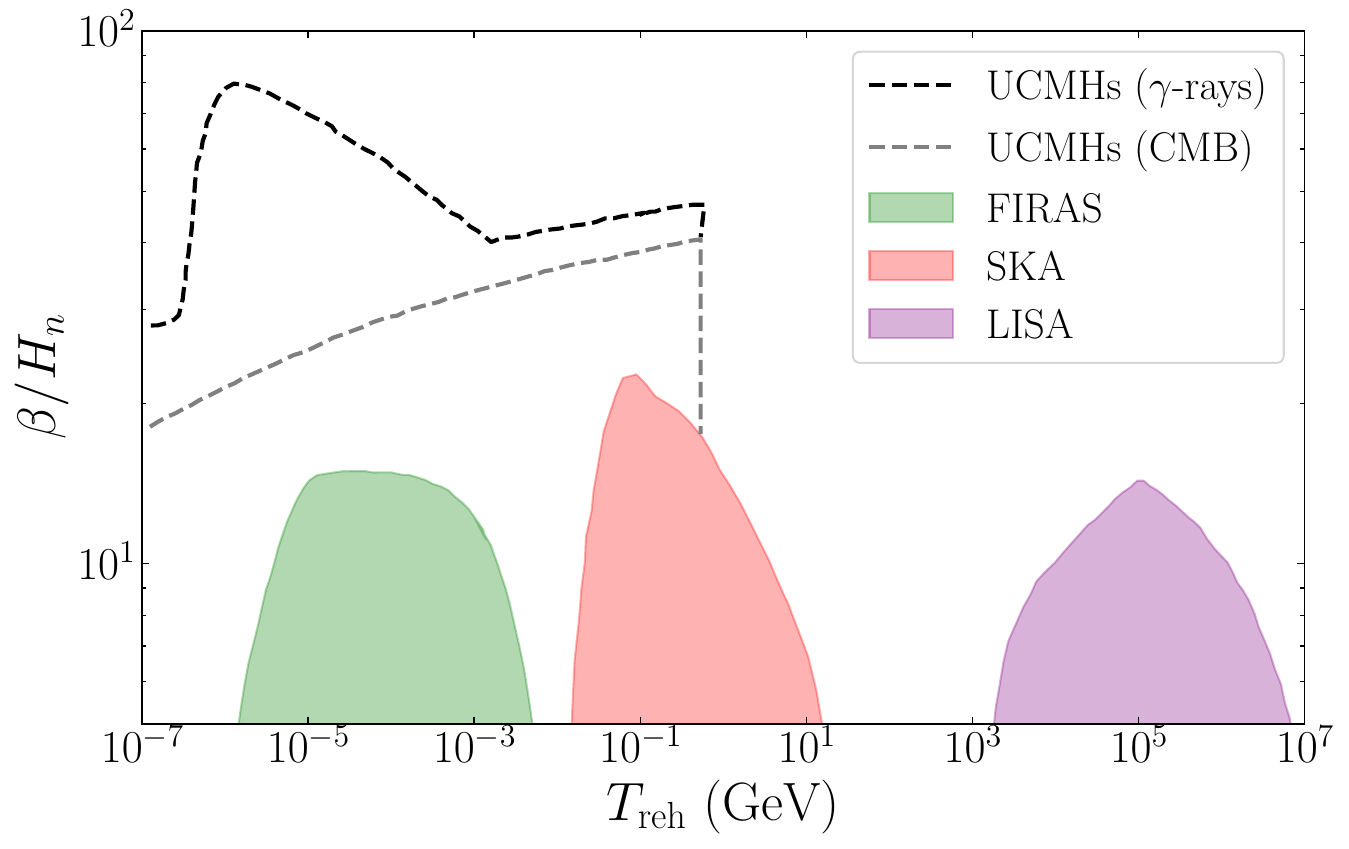}
    \caption{ Similar to Fig.~\ref{fig:cons}, but translates various excluded regions into $T_{\rm reh}-\beta/H_{n}$ plane using a monochromatic power spectrum $P_{\mathcal{R}}(k= k_{\rm max})$. The two UCMH lines represent excluded regions below.}
    \label{fig:beta_T}
\end{figure}

In Fig.~\ref{fig:cons}, we present these existing and forecast constraints together with selected benchmark $P_{\mathcal{R}}(k)$s for various $\beta/H_{n}$ and $T_{\rm reh}$. In Fig.~\ref{fig:beta_T}, we further map the excluded regions on the $T_{\rm reh}-\beta/H_{n}$ plane with a monochromatic power spectrum $P_{\mathcal{R}}(k= k_{\rm max})$, where we have truncated the lower limit of $\beta/H_{n}$ to $4$ to ensure the validity of the fitting formula in Eq.~(\ref{eq:pr}). However, since these constraints are obtained from a spiky power-spectrum enhancement, their applicability to $P_{\mathcal{R}}(k)$ sourced by FOPTs requires further investigation. Moreover, the constraints from UCMHs~\cite{Delos:2018ueo,FrancoAbellan:2023sby} are based on the assumption that dark matter annihilates into $b\bar{b}$ with a mass $m_{\rm DM}=1$ TeV and $s$-wave thermal relic cross section $\langle \sigma v \rangle_0=3 \times 10^{-26} \mathrm{cm}^{3}\mathrm{s}^{-1}$. Otherwise, the corresponding limits would be significantly relaxed. We also note that constraints from UCMHs based on extended dark objects are available~\cite{Bringmann:2025cht}. These constraints can be derived from CMB accretion, wide binary evaporation, or lensing observations of the “Icarus” star, without assuming dark matter annihilation. They typically place constraints on $P_{\mathcal{R}}(k)$ comparable in range and magnitude to the CMB energy injection constraints (gray dashed lines in Fig.~\ref{fig:cons}).


\subsection{Scalar-induced gravitational waves}

The SIGWs produced by these curvature perturbations can represent a significant secondary source of phase transition gravitational waves~\cite{Lewicki:2024ghw}, distinct from the primary gravitational waves produced by the dynamics of phase transitions, such as bubble collisions~\cite{Huber:2008hg}, sound waves~\cite{Hindmarsh:2013xza}, or turbulence~\cite{Caprini:2009yp}, et. al. 

The spectrum of the SIGW can be computed from
\begin{align}
\Omega_{\mathrm{SIGW}}h^{2} 
&=\Omega_{r,0}h^2 \frac{g_\rho(T_{\rm{reh}})}{g_\rho(T_0)}\left(\frac{g_s(T_{\rm{reh}})}{g_s(T_0)}\right)^{-4/3}\Omega_{\mathrm{SIGW}}(T_{\rm{reh}}) \nonumber\\
&=1.65\times10^{-5}\left(\frac{\Omega_{r,0}h^2}{4.18\times10^{-5}}\right)\left(\frac{g_\rho(T_{\rm{reh}})}{100}\right) \nonumber\\
&\quad \times\left(\frac{g_s(T_{\rm{reh}})}{100}\right)^{-4/3} \Omega_{\mathrm{SIGW}}(T_{\rm{reh}})\,,
\end{align}
where $\Omega_{r,0}h^2$ is the energy density ratio of radiation today. The quantity $\Omega_{\mathrm{SIGW}}(T_{\rm{reh}})$ characterizes the induced gravitational waves at the reheating temperature $T_{\rm reh}$~\cite{Kohri:2018awv,Espinosa:2018eve}, 
\begin{align}
\Omega_{\mathrm{SIGW}}(T_{\rm{reh}}) = \int_{0}^{\infty} \mathrm{d}v \int_{|1-v|}^{1+v} \mathrm{d}u\, T_{\rm{RD}}(u,v) P_{\mathcal{R}}(ku) P_{\mathcal{R}}(kv) \,,
\end{align}
and the transfer function $T_{\rm{RD}}$ is defined as
\begin{align}
&T_{\rm{RD}}(u,v) = \left(\frac{4v^{2} - (1 - u^{2} + v^{2})^{2}}{4u^{2}v^{2}}\right)^{2} y^{2} \nonumber\\
&\times  \left\{\frac{\pi^{2}}{4} y^{2} \Theta[u + v - c_r^{-1}] + \left(1 - \frac{1}{2} y \ln \left|\frac{1 + y}{1 - y}\right|\right)^{2}\right\} \,,
\end{align}
with $y =(u^{2} + v^{2} - c_r^{-2})/(2uv)$.

\begin{figure}[t!]
    \centering
    \includegraphics[width=0.95\linewidth]{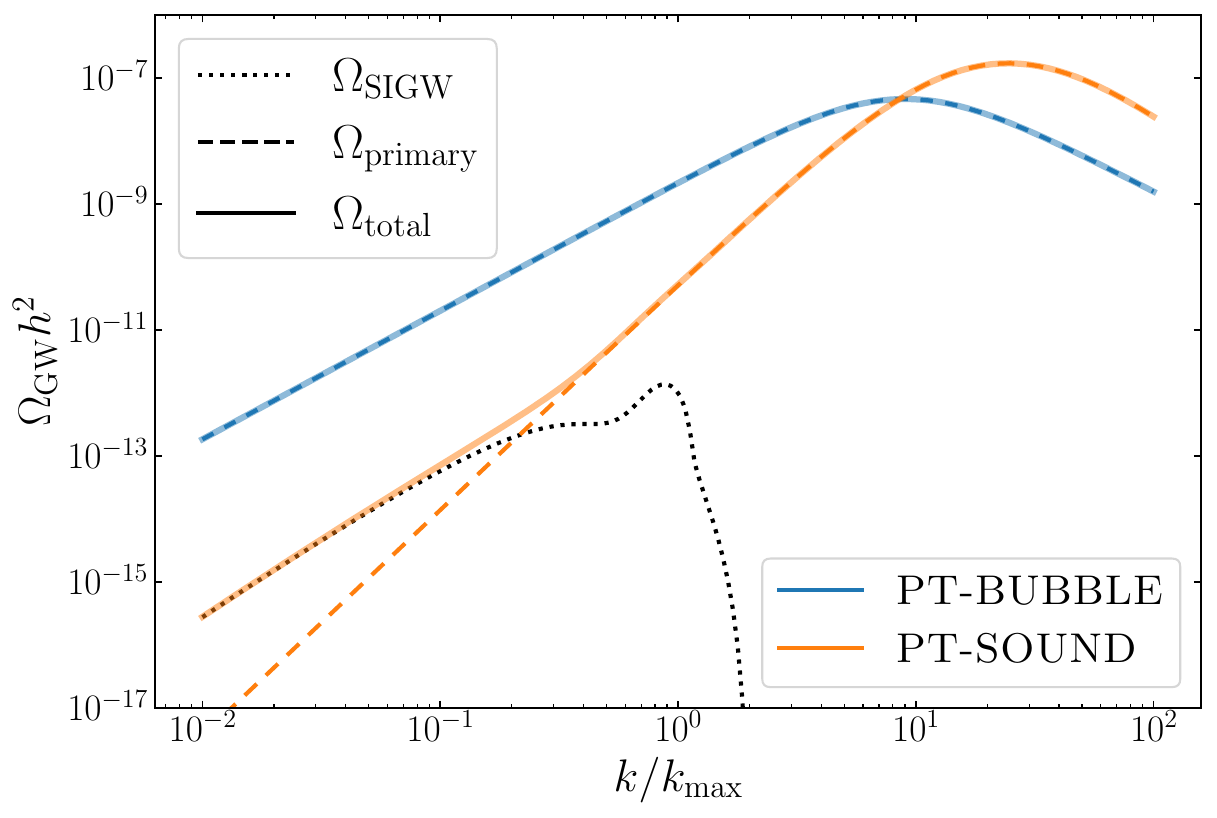}
    \caption{GWs spectra from \textsc{PT-BUBBLE} and \textsc{PT-SOUND} models as the primary GW sources together with SIGWs from curvature perturbations. }
    \label{fig:gw}
\end{figure}

\begin{figure*}[htbp]
    \centering
    \includegraphics[width=0.47\linewidth]{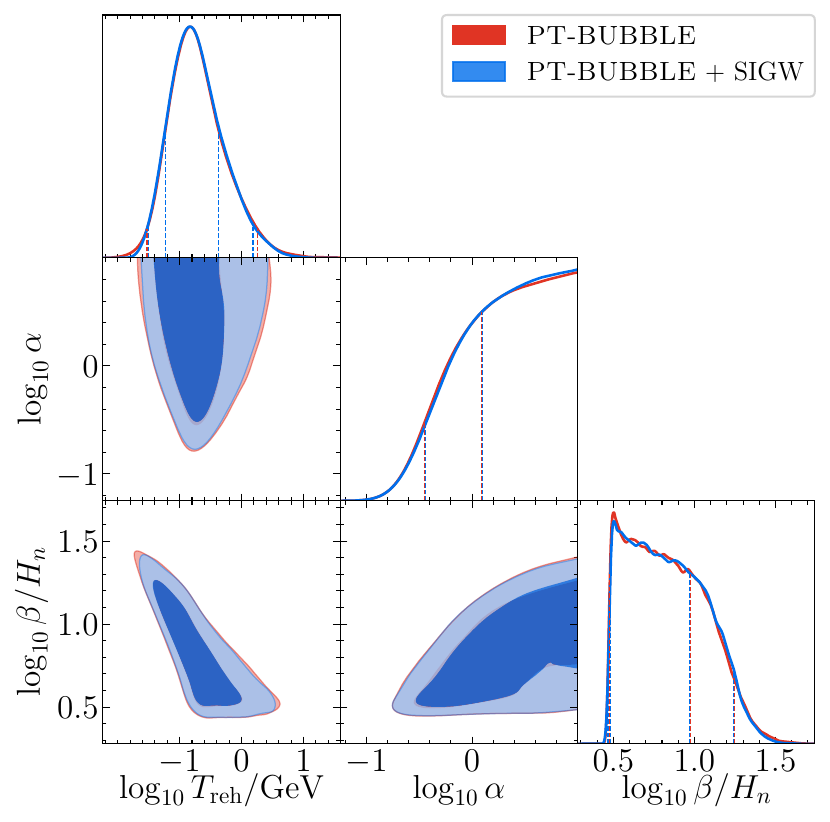}
    \includegraphics[width=0.47\linewidth]{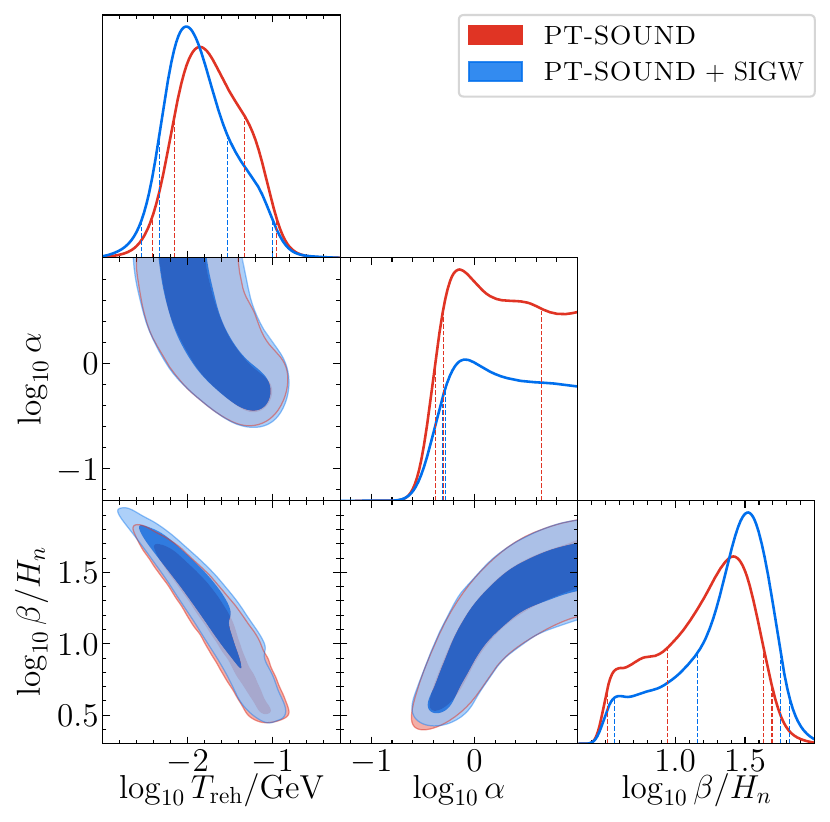}
    \caption{Posterior distributions of phase transition parameters from NANOGrav-15 data, based on \textsc{PT-BUBBLE} (left) and \textsc{PT-SOUND} (right) models, with and without considering SIGW introduced by curvature perturbations. The contours show both the 68\% (darker) and 95\% (lighter) Bayesian credible regions. And the vertical lines in the 1D marginalized distributions also represent 68\% and 95\% Bayesian credible intervals.}
    \label{fig:3}
\end{figure*}

The total observed GW spectrum is then composed of a primary component and a secondary component from SIGWs:
\begin{align}
\Omega_{\mathrm{total}}h^{2}=\Omega_{\mathrm{primary}}h^{2}+\Omega_{\mathrm{SIGW}}h^{2}\,.
\end{align}
As an example, we investigate two typical FOPT models: bubble collisions (\textsc{PT-BUBBLE}) and sound waves (\textsc{PT-SOUND}). The fitting formulas for the GWs in both models, detailed in~\cite{NANOGrav:2023hvm}, share the same broken power-law shape function:
\begin{align}
S(x)=\frac{(a+b)^c}{\left(bx^{-a/c}+ax^{b/c}\right)^c},
\end{align}
where $a$, $b$, and $c$ are shape parameters. In Fig.~\ref{fig:gw}, we present both the primary and secondary GW spectra, calculated with phase transition parameters $\alpha=10$, $\beta / H_n = 7$, and $T_{\rm reh} = 0.05\;\rm GeV$. The shape parameters are set to their best-fit values from the NANOGrav 15-year dataset~\cite{NANOGrav:2023hvm}: $a=2.04$, $b=1.97$, $c=2.03$ for \textsc{PT-BUBBLE}, and $a=3.58$, $b=2.87$, $c=4.16$ for \textsc{PT-SOUND}. Fig.~\ref{fig:gw} clearly illustrates that the SIGW spectrum peaks at a larger scale compared to the primary one. However, our gauge-invariant calculation suggests that this potential secondary SGWB source is likely subdominant compared to the primary source. Including this secondary component may only introduce marginal modifications at low frequencies for the \textsc{PT-SOUND} model.

To further investigate the impact of SIGWs on parameter estimation, we analyze NANOGrav 15-year data and illustrate the modifications introduced by this secondary source. Our analysis focuses on three phase transition parameters: the strength $\alpha$, the reheating temperature $T_{\rm reh}$, and the nucleation rate $\beta/H_n$. 
We employ the \texttt{PTArcade} code~\cite{Mitridate:2023oar,mitridate_2023} to derive posterior distributions for these model parameters. The results for the \textsc{PT-BUBBLE} and \textsc{PT-SOUND} models are presented in Fig.~\ref{fig:3}. We observe negligible modifications to the posterior distributions for the \textsc{PT-BUBBLE} model and only marginal changes for the \textsc{PT-SOUND} model. The small alterations of the \textsc{PT-SOUND} result from the low-frequency modifications introduced by the SIGW. Overall, we conclude that the inclusion of GWs induced by super-horizon curvature perturbations does not significantly affect the interpretation of the FOPT models in the context of current PTA data.

\section{Conclusions}

Slow and strongly supercooled FOPTs, which are favored by recent PTA data, can generate super-horizon inhomogeneities.
In this study, we performed a gauge-invariant calculation of the curvature perturbations induced by these inhomogeneities through constructing gauge-invariant entropy perturbations.
Our results indicate that the inhomogeneities derived from separate universe simulations cannot be interpreted as perturbations within the comoving gauge, a choice commonly employed to assess the abundance PBHs. Furthermore, the distribution of the density contrast, estimated from the gauge-invariant comoving curvature perturbations, exhibits a positive tail that remains significantly below the conventional PBH formation threshold. Consequently, we conclude that the formation of PBHs through this mechanism is highly unlikely.

We also present an empirical template for these super-horizon curvature perturbations, which can be utilized to compute potential cosmological observables. As a demonstration, 
we examine the consistency of $P_{\mathcal{R}}$ generated by FOPTs with current and future constraints on primordial curvature perturbations. Our results indicate that slow FOPTs may be constrained by future SKA data, whereas relatively fast FOPTs are subject to constraints derived from UCMH observations.
We also calculate the SIGWs generated by these perturbations, representing a potential secondary GW source. However, owing to the suppressed amplitude of the curvature perturbations, this secondary GW source has only a marginal impact on conventional phase transition models constrained by the NANOGrav 15-year data. We further note that for super-cooled and slow FOPTs, more refined models incorporating the effects of cosmic expansion may be required~\cite{Guo:2020grp,Lewicki:2025hxg,Yamada:2025cfr,Yamada:2025hfs}. A detailed investigation of this issue is deferred to future work.

In summary, this study proposes a new framework for quantifying the large-scale gravitational effects of FOPTs as curvature perturbations. Our gauge-invariant analysis disfavors both the formation of PBHs and a significant secondary GW source by SIGW via this mechanism. The template for the comoving curvature perturbations presented in this work can be applied to future investigations of other potential observables originating from the large-scale inhomogeneities associated with FOPTs.

\begin{acknowledgments}
The authors thank Ryusuke Jinno, Shaojiang Wang and Ke-pan Xie for fruitful discussions. X.W. and C.B. are supported by Australian Research Council grants DP220100643 and LE250100010. R.D. is supported in part by the National Key R\&D Program of China (No. 2021YFC2203100). 
C.T. is supported by the National Natural Science Foundation of China (Grants No. 12405048) and the Natural Science Foundation of Anhui Province (Grants No. 2308085QA34). 

\end{acknowledgments}

\bibliography{ref.bib}

\end{document}